\documentclass[aps,pre,twocolumn,showpacs,amsmath,amssymb,final,letterpaper]{revtex4}
\usepackage{graphicx}
\usepackage{subfigure}
\usepackage{txfonts}

\begin{document}

\title{Structural preferential attachment: \\ Stochastic process for the growth of scale-free, modular and self-similar systems}

\author{Laurent H\'ebert-Dufresne}
\author{Antoine Allard}
\author{Vincent Marceau}
\author{Pierre-Andr{\'e} No\"el}
\author{Louis J. Dub\'e}

\affiliation{D\'epartement de Physique, de G\'enie Physique, et d'Optique, Universit\'e Laval, Qu\'ebec (Qu{\'e}bec), Canada G1V 0A6}
\date{\today}
\begin{abstract}
Many complex systems have been shown to share universal properties of organization, such as \emph{scale independence}, \emph{modularity} and \emph{self-similarity}. We borrow tools from statistical physics in order to study \emph{structural preferential attachment} (SPA), a recently proposed growth principle for the emergence of the aforementioned properties. We study the corresponding stochastic process in terms of its time evolution, its asymptotic behavior and the scaling properties of its statistical steady state. Moreover, approximations are introduced to facilitate the modelling of real systems, mainly complex networks, using SPA. Finally, we investigate a particular behavior observed in the stochastic process, the \emph{peloton dynamics}, and show how it predicts some features of real growing systems using prose samples as an example.
\end{abstract}
\pacs{89.75.Da, 89.75.Fb, 89.75.Hc, 89.75.Kd, 89.65.Ef}
\maketitle

\section{Introduction}

In a recent contribution, we have proposed a model of network organization \cite{SPA} based on a generalization of the classical preferential attachment principle (PA) \cite{Simon55, barabasi99} to a higher order: structural preferential attachment (SPA). In this model, elements of the system join and create structures. In all attachment events, both the element and the structure involved are chosen proportionally to their past activities. Elements can represent money being invested, written words, individuals in a social network, proteins or websites, while the structures can be business firms, semantic fields, friendships and communities, protein complexes or types of activities and interest \cite{Simon55, barabasi99, palla05, ahn}.

\begin{figure}[b!]
  \centering
  \includegraphics[width=0.875\columnwidth]{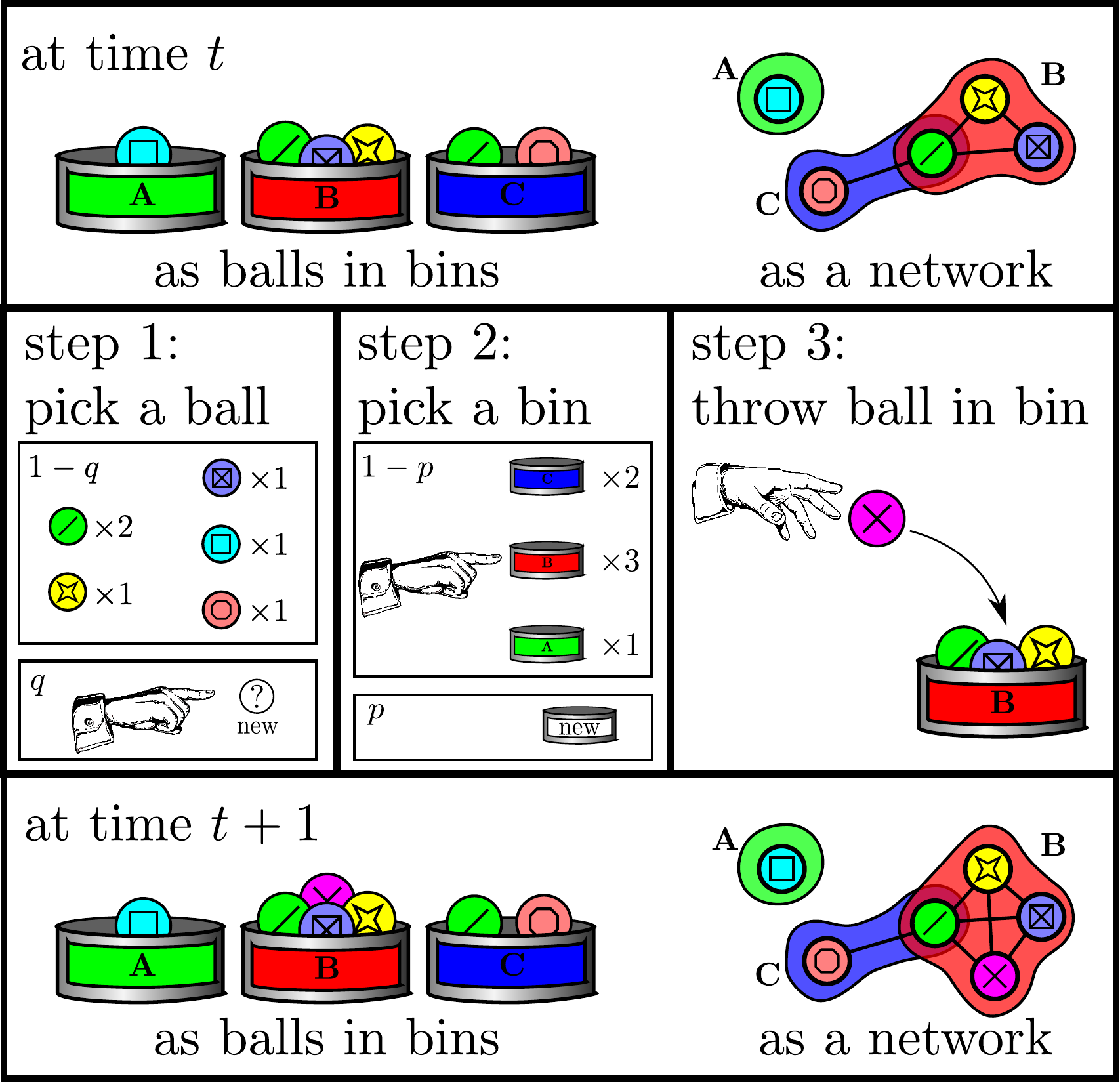}
  \caption{(color online). A step of node-based SPA.}
  \label{SPA_scheme}
\end{figure}

SPA can be described by the following stochastic process (see Fig. \ref{SPA_scheme} for a visual aid). At every time step, an element joins a structure. With probability $q$, the element is a new one; or with probability $1-q$, it is chosen among existing elements proportionally to the current number of structures to which they belong (i.e., their \emph{membership} number). Moreover, with probability $p$, the structure is a new one of size $s$; or with probability $1-p$, it is chosen among existing structures proportionally to the current number of elements they possess (i.e., their \emph{size}). Whenever the structure is a new one, the remaining $s-1$ elements involved in its creation are once again preferentially chosen among existing nodes. The basic structure size $s$ is called the \emph{system base} and refers to the smallest structural unit of the system. For example, if $s=1$, the system base is simply the elements themselves and we refer to this version as \emph{node-based} SPA, while if $s=2$, the system base is a pair of elements resulting in \emph{link-based} SPA.

This stochastic process can either be seen as a scheme of throwing balls (the elements) in bins (the structures) or as a process of network growth. In the latter, the elements are the \emph{nodes} of the network while the structures represent significant topological patterns, motifs, modules or \emph{communities}, within which elements are linked.

SPA results in the growth of \emph{modular} systems, because modules (or structures) are the basic building blocks of the model. These systems are also \emph{scale-free}, in the sense that their main statistical features (membership and size distributions) converge toward power laws (free of any characteristic scale) as a result of the preferential attachment principle \cite{Simon55, barabasi99}. Finally, these systems are said to be \emph{self-similar} as different levels of organization follow the same general behavior: elements are interconnected with one another by sharing structures in the same way the structures themselves are interconnected by sharing elements.

In this paper, we borrow tools from statistical physics to study SPA in detail. In Sec. \ref{sec:process}, an exact description of SPA is obtained by writing the corresponding discrete stochastic process. From this description, we obtain the statistical steady-state of the resulting system with asymptotic expressions for its scaling behaviors. In Sec. \ref{sec:approximations}, some useful approximations are introduced and studied in order to facilitate the comparison between systems produced by SPA and real-world systems, using the \textit{cond-mat arXiv} co-author network as an example. In order to investigate the validity of these approximations, we then study the existence of correlations between elements and structures, in both the SPA process and in the \textit{cond-mat arXiv}. Lastly, in Sec. \ref{sec:peloton}, we highlight an interesting behavior of discrete PA processes, which we call the \emph{peloton dynamics}, by comparing the initial stochastic process with an explicit solution for the time evolution of the continuous time version (further details are presented in the Appendices A and B). We then seek empirical evidences of this behavior in growing prose samples. A conclusion summarizes our results.

\section{Stochastic process \label{sec:process}}

\subsection{Time evolution}
To follow the growth of a system as prescribed by the SPA process, we separate elements and structures. We distinguish nodes by their respective number of memberships, $m$, and structures by their respective size, $n$, as these are the only features relevant to their evolution. Let $\tilde{N}_m(t)$ be the mean number of elements (or \emph{nodes} to use the network terminology) with $m$ memberships and $\tilde{S}_{\! n}(t)$ be the mean number of structures  of size $n$. Throughout the paper, tildes are used in quantities describing absolute numbers. Also note that as we follow the mean distribution of these quantities, we restrict ourselves to a deterministic approximation of the process.

At each time step, the evolution of these quantities is twofold: first, a constant increment for potential new nodes and structures; second, an operation corresponding to the preferential growth of existing nodes and structures. More clearly, each time step corresponds to an iteration of the following rule:
\begin{align}
\tilde{N}_m(t+1) = & \tilde{N}_m(t) + q\delta _{m1} \nonumber \\
& + \frac{1\! -\! q\! +\! p\left(s\! -\! 1\right)}{t\left[1+p\left(s\! -\! 1\right)\right]}\left[\left(m\! -\! 1\right)\tilde{N}_{m\! -\! 1}(t) -\!  m\tilde{N}_m(t)\right] \label{d1} \\
\tilde{S}_{\! n}(t+1) = & \tilde{S}_{\! n}(t) + p\delta _{ns} \nonumber \\
 & + \frac{1-p}{t\left[1+p\left(s\! -\! 1\right)\right]}\left[\left(n\! -\! 1\right)\tilde{S}_{\! n\! -\! 1}(t) -\! n\tilde{S}_{\! n}(t)\right] \; . \label{d2}
\end{align}
The two increments $q\delta _{m1}$ and $p\delta _{ns}$, where $\delta _{ij}$ is the Kronecker delta, correspond to birth events for elements (with one membership) and structures (of size $s$), respectively. The last increments correspond to the growth of old entities, where a compartment has a negative effect on itself and a positive effect on its neighboring compartment (e.g., $\tilde{N}_m \rightarrow \tilde{N}_{m+1}$) at a given rate and the denominator $t\left[1+p(s-1)\right]$ normalizes the preferential attachment probabilities.

This iterative description is straightforward, yet we can define the system in closed form by using generating functions (GFs) \cite{wilf}. We define two functions whose power series coefficients correspond to the elements of our two ensembles:
\begin{equation}
\widetilde{\mathcal{N}}(x; t) = \sum _m \tilde{N}_m(t) x^m \;\;\; \textrm{and} \;\;\; \widetilde{\mathcal{S}}(x; t) = \sum _n \tilde{S}_{\! n}(t)x^n
\end{equation}
In terms of these GFs, Eqs. (\ref{d1}) and (\ref{d2}) can be rewritten as:
\begin{eqnarray}
\widetilde{\mathcal{N}}(x;t+1) & = & \left(1 + \dfrac{\Gamma _s}{t}x\left(x-1\right)\frac{d}{d x}\right) \widetilde{\mathcal{N}}(x;t) + qx \; ; \label{abs1}\\
\widetilde{\mathcal{S}}(x;t+1) & = & \left(1 + \dfrac{\Omega _s}{t}x\left(x-1\right)\frac{d}{d x}\right) \widetilde{\mathcal{S}}(x;t) + px^s \; , \label{abs2}
\end{eqnarray}
where we have also introduced
\begin{equation}
\Gamma _s = \frac{1-q+p(s-1)}{1+p(s-1)} \;\;\; \textrm{and } \;\; \Omega _s = \frac{1-p}{1+p(s-1)} \; .
\label{gam_ome}
\end{equation}

A similar description can be obtained in terms of the corresponding probability generating functions (PGFs), $\mathcal{N}(x;t)$ and $\mathcal{S}(x;t)$, which generate the distributions of memberships per element and size per structures respectively. To transform the previous description in terms of these PGFs, note that the mean numbers of elements, $\tilde{N}_m$, or structures, $\tilde{S}_{\! n}$, in a given state corresponds to the proportion of such elements, $N_m$, or structures, $S_{\! n}$, multiplied by the mean total number of elements, $qt$, or structures, $pt$, expected at time $t$. One can now rewrite Eqs. (\ref{abs1}) and (\ref{abs2}) in terms of $\mathcal{N}(x;t)$ and $\mathcal{S}(x;t)$ by multiplying these functions by $qt$ and $pt$, respectively:
\begin{eqnarray}
\left(t+1\right)\mathcal{N}(x;t+1) & = & \left(t + \Gamma _s x\left(x-1\right)\frac{d}{d x}\right) \mathcal{N}(x;t) + x  \label{rel1}\\
\left(t+1\right)\mathcal{S}(x;t+1) & = & \left(t + \Omega _s x\left(x-1\right)\frac{d}{d x}\right) \mathcal{S}(x;t) + x^s . \label{rel2}
\end{eqnarray}

As we will see in what follows, the description in terms of PGFs is generally more useful and will hereafter be used in our results to validate the analytical description.

\begin{figure*}[]
  \centering
  \subfigure[]{\includegraphics[width=0.32\linewidth]{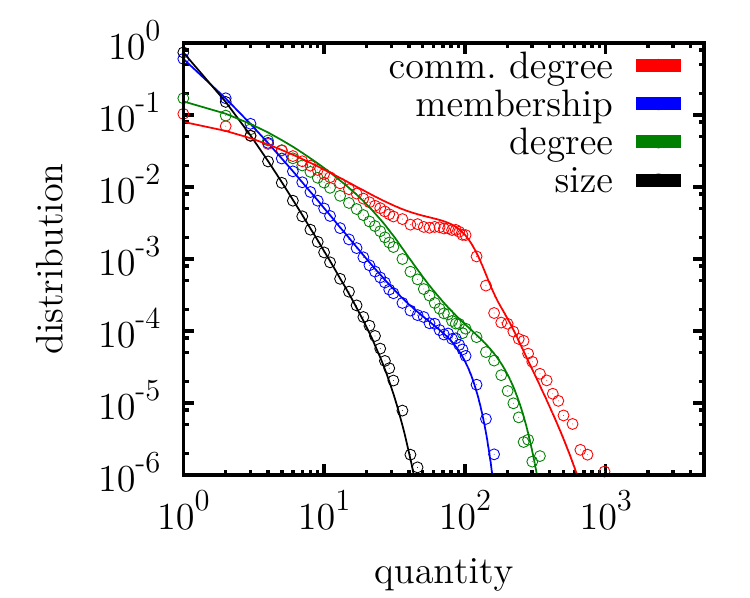} \label{evo250}}         
  \subfigure[]{\includegraphics[width=0.32\linewidth]{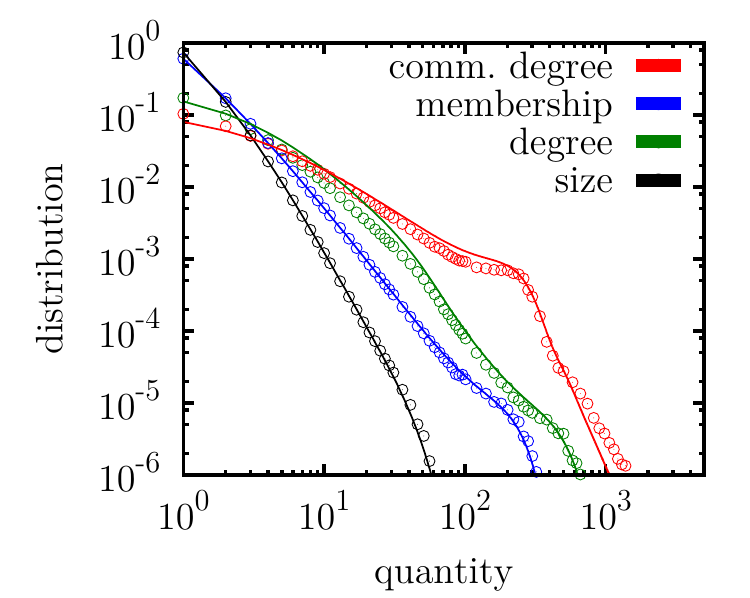} \label{evo1000}}
  \subfigure[]{\includegraphics[width=0.32\linewidth]{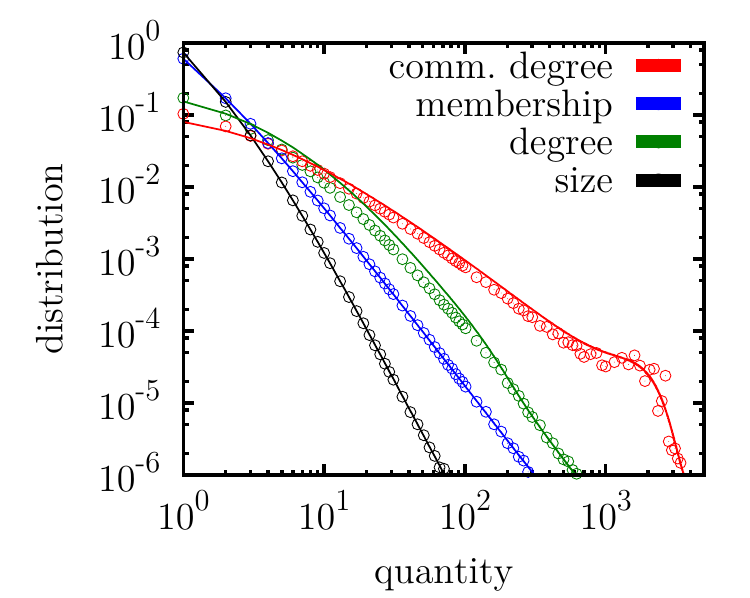} \label{evo25000}}
  \caption{(color online). Time evolution of node-based SPA ($s=1$) using $q=0.35$ and $p=0.65$ for the four main characteristics of the topology: memberships, community size, node degree and community degree. Snapshots are taken when the systems reach (a) 250 structures, (b) 1000 structures and (c) 25 000 structures. Shown by markers are Monte Carlo results averaged over 25 000 simulations; analytical predictions of Eqs. (\ref{rel1}) - (\ref{comm_deg}) are shown with continuous lines.}
  \label{timeevo}
\end{figure*}

\begin{figure*}[]
  \centering
  \subfigure[]{\includegraphics[trim = 3mm 0mm 0mm 0mm, clip, width=0.24\linewidth]{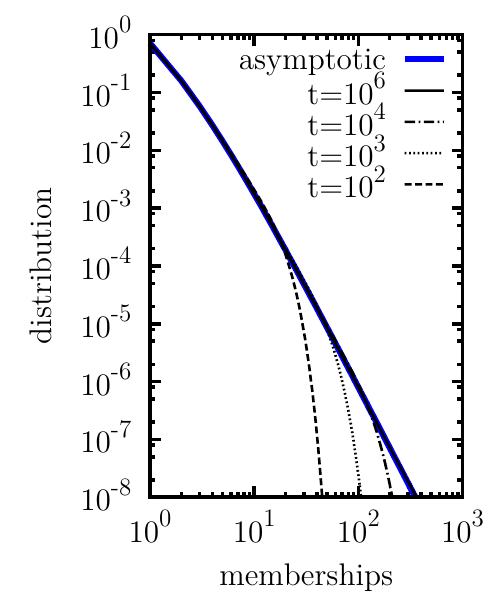} \label{asympto1}}         
  \subfigure[]{\includegraphics[trim = 3mm 0mm 0mm 0mm, clip, width=0.24\linewidth]{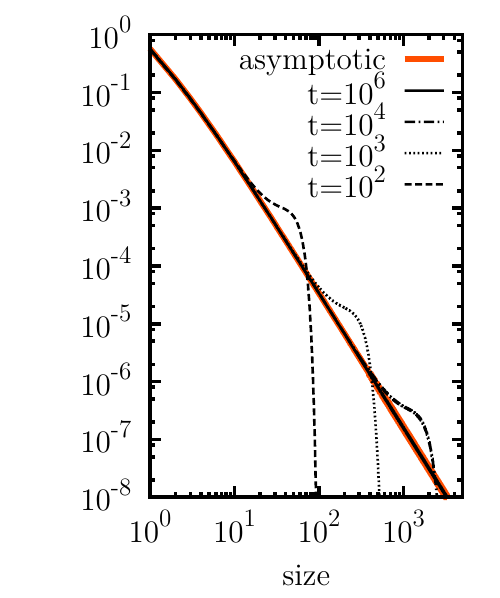} \label{asympto2}}
  \subfigure[]{\includegraphics[trim = 3mm 0mm 0mm 0mm, clip, width=0.24\linewidth]{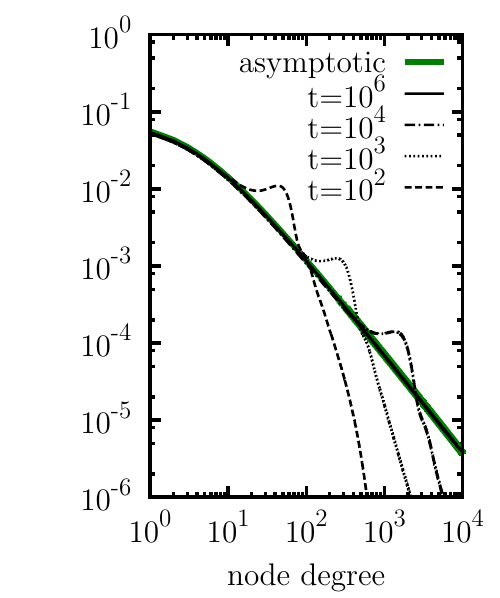} \label{asympto3}}
  \subfigure[]{\includegraphics[trim = 3mm 0mm 0mm 0mm, clip, width=0.24\linewidth]{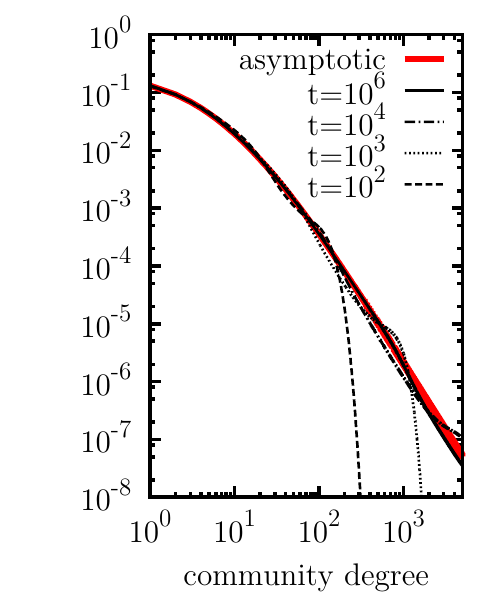} \label{asympto4}}
  \caption{(color online). Convergence of the time evolution governed by  Eqs. (\ref{rel1}) - (\ref{comm_deg}) toward the equilibrium predicted by Eq. \ref{powers} for (a) the membership distribution, (b) the size distribution, (c) node degree distribution and (d) community degree distribution in node-based SPA ($s=1$) using $q=0.6$ and $p=0.25$.}
  \label{asympto}
\end{figure*}

\subsection{Degree distributions \label{degdis}}

PGFs provide simple ways to evaluate secondary properties of a given state. For example, the node degree distribution and the community degree distribution. The former describes how many elements can be reached from a randomly chosen element, in other words, the number of links connected to this node in the network representation. The latter refers to a similar concept, namely, the number of structures that overlap (by sharing elements) with one randomly chosen structure. 

To illustrate how this calculation is performed, one can simply refer to the composition property of PGFs. We first pick a random element whose membership distribution is generated by $\mathcal{N}(x;t)$. For every possible value of its membership number $m$, we sum over all possible cases for the different sizes of these structures. However, we know that all of these $m$ structures have \emph{at least} one element. It is thus $k$ times more likely that one of these $m$ structures is a structure of size $k$ than a structure of size one. Furthermore, we do not want to count the initial element we chose, and will thus reduce the size of each structure by one. Hence, their size distribution is not generated by $\mathcal{S}(x;t)$, but instead by $\mathcal{S}'(x;t) /\mathcal{S}'(1;t)$, where the denominator acts as a normalisation factor. Knowing that the convolution of two sequences is generated by the product of the corresponding PGFs, one can take the $m$-th power of the new size PGF to obtain the PGF for the sum of $m$ structures. Finally, we sum over all possible values of $m$ to obtain \cite{newman03}:
\begin{align}
D(x;t) & = \sum _m N_m \left[ \mathcal{S}'(x;t) /\mathcal{S}'(1;t) \right]^m \nonumber \\
& = \mathcal{N}\left(\left[\mathcal{S}'(x;t)\bigg /\mathcal{S}'(1;t)\right], t\right) \; .
\label{node_deg}
\end{align}
Using the same logic for structures and their community degree, one can write:
\begin{equation}
C(x;t)  = \mathcal{S}\left(\left[\mathcal{N}'(x;t)\bigg /\mathcal{N}'(1;t)\right], t\right) \; .
\label{comm_deg}
\end{equation}
%The last two PGFs are denoted as $P_n(x,t)$ and $P_c(x,t)$ because they generate the \emph{potential node degree distribution} and the \emph{potential community degree distribution} at time $t$. They are potential distributions because two elements can share more than one structure, just as two structures can overlap through more than one element; but these cases only contribute to a single degree. These multiple links with the same entity are a consequence of the finite size of the system, because their probabilities of occurrences would be null in the infinite limit. To take the size of the system into account, we must answer the following question: what is the probability that with $i$ potential degree, are shared with only $j$ element?

The self-similarity between different levels of organization in the systems created by SPA stems from the similarity between Eqs. (\ref{node_deg}) and (\ref{comm_deg}). As long as $\mathcal{N}(x;t)$ and $\mathcal{S}(x;t)$ are similar, the various possible compositions, which represent different organization properties, will also be similar.

The validation of our analytical description for the time evolution of SPA is presented on Fig. \ref{timeevo} using Monte Carlo simulations. The initial conditions of all systems (i.e., the state of the system at $t=0$), in both numerical simulation and analytical integration, consist of a single structure containing a single element; this remains true throughout the paper. Note that our calculations for the degree distributions are merely approximations because they suppose \emph{homogeneous mixing} between elements and structures, while an element with $m=i$ might not see exactly the same size distribution as an element with $m=j$. Such element-structure correlations are investigated in Sec. \ref{sec:correlations}.

\subsection{Statistical equilibrium}

\begin{figure}
  \centering
  \includegraphics[width=0.35\textwidth]{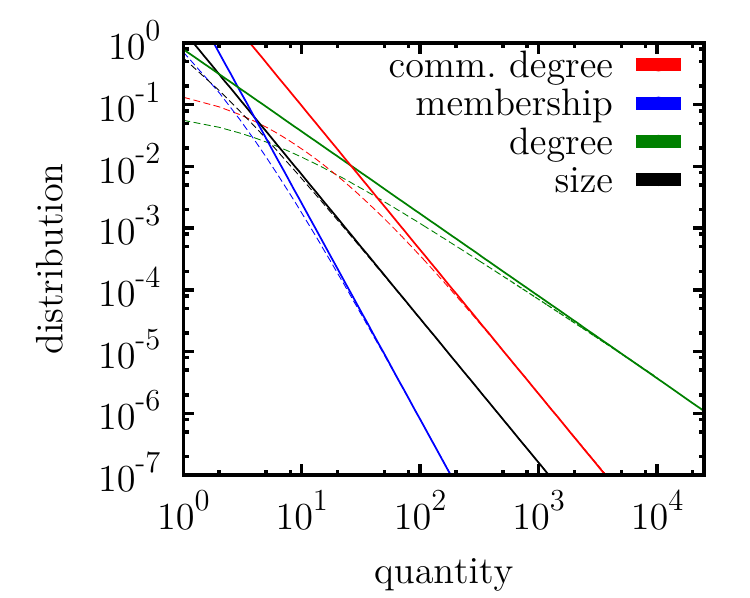}
  \caption{(color online). Validation of Eqs. (\ref{scale1}) and (\ref{scale2}) as predictions for the asymptotic scaling behaviors of the main statistical distributions (dashed lines: steady-state solutions, continuous line: scaling predictions) for node-based SPA ($s=1$) using $q=0.6$ ($\gamma _N = 7/2$) and $p=0.25$ ($\gamma _S = 7/3$).}
  \label{scaling}
\end{figure}

The statistical equilibrium can be imposed by setting $\mathcal{N}(x;t+1) = \mathcal{N}(x,t) \equiv \mathcal{N}^*(x)$ and $\mathcal{S}(x,t+1) = \mathcal{S}(x,t) \equiv \mathcal{S}^*(x)$ in Eqs. (\ref{rel1}) and (\ref{rel2}), yielding:
\begin{eqnarray}
\mathcal{N}^*(x) & = & \Gamma _s x\left(x-1\right)\frac{d}{d x}\mathcal{N}^*(x) + x \; ; \label{eq1}\\
\mathcal{S}^*(x) & = & \Omega _s x\left(x-1\right)\frac{d}{d x}\mathcal{S}^*(x) + x^s \; . \label{eq2}
\end{eqnarray}
These ordinary differential equations can be solved straightforwardly to obtain their solutions in terms of hypergeometric functions of the form ${}_2F_1\left(a,b;c;x\right)$:
\begin{equation}
\mathcal{N}^*(x) = \frac{x}{1+\Gamma _s} \; {}_2F_1\left(1,1;2+\frac{1}{\Gamma _s};x\right) \; ,
\end{equation}
and:
\begin{equation}
\mathcal{S}^*(x) = \frac{(s-1)!\Omega _s^{s-1}x^s}{1+s\Omega _s} \; {}_2F_1\left(1,s;(s+1)+\frac{1}{\Omega _s};x\right) \; .
\end{equation}
The statistical equilibrium for the two distributions of interest can now be obtained through the power series coefficients of these two functions:
\begin{eqnarray}
& N_m^*(s) = \dfrac{\prod _{k=1}^{m-1} k\Gamma _s}{\prod _{k=1}^{m} \left(1+k\Gamma _s\right)} \;, &\; S_n^*(s) = \dfrac{\prod _{k=s}^{n-1} k\Omega _s}{\prod _{k=s}^{n} \left(1+k\Omega _s\right)} \; .
\label{powers}
\end{eqnarray}
These solutions for the asymptotic behavior of the statistical distributions can be validated through comparison with the long term behavior of our predicted time evolution, as done in Fig. \ref{asympto}.

\begin{figure*}[]
  \centering
  \subfigure[]{\includegraphics[width=0.32\linewidth]{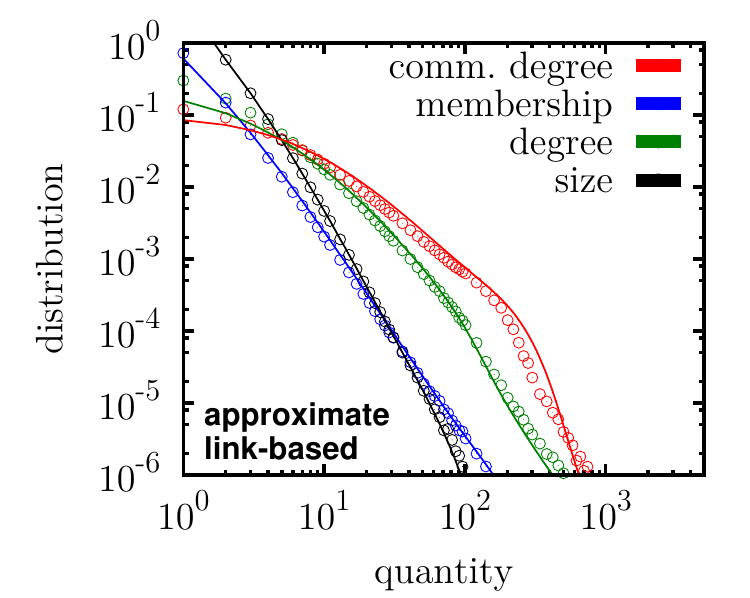} \label{bases1}}
  \subfigure[]{\includegraphics[width=0.32\linewidth]{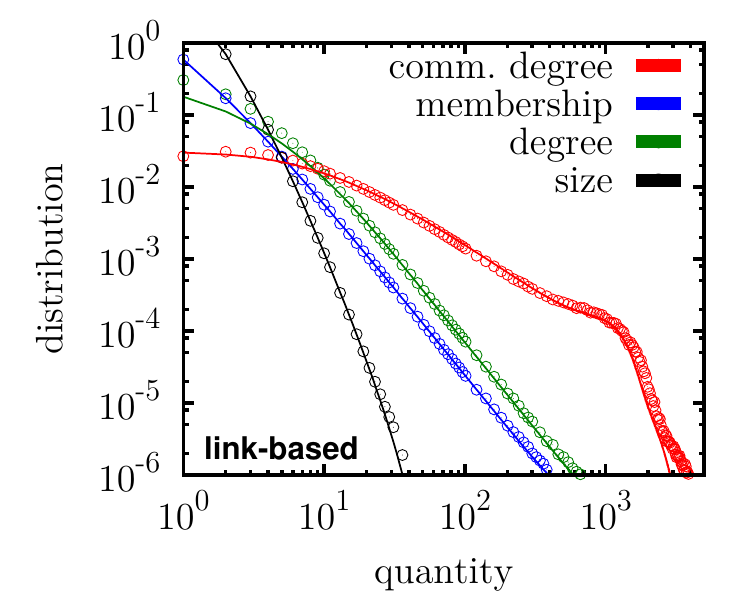} \label{bases2}}
  \subfigure[]{\includegraphics[width=0.32\textwidth]{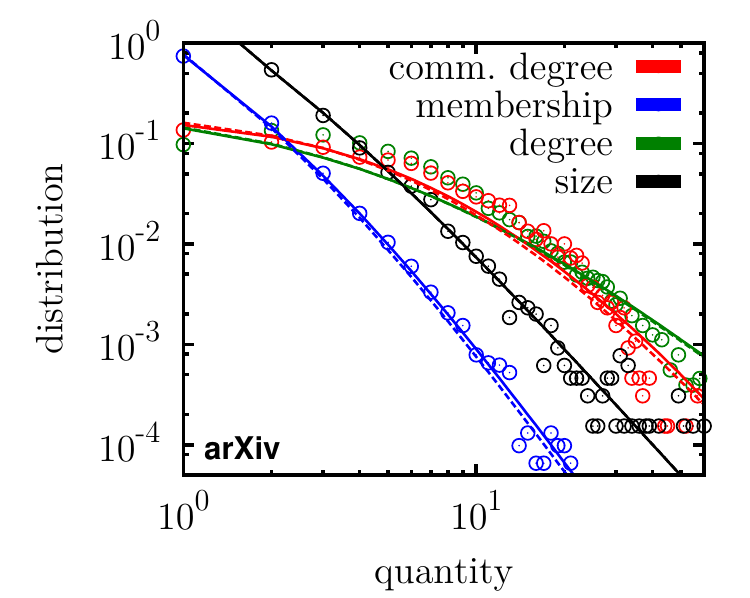} \label{arXiv}}
  \caption{(color online). (a) and (b) Analytical predictions (lines) and simulations (markers) for a) an approximation of link-based system using a node-based SPA process and b) the link-based SPA using the same parameters as in (a). (c) Community structure of the \emph{cond-mat arXiv} as measured by a link community algorithm \cite{ahn} (dots) and as modelled by a link-based SPA ($q_2 =0.95$; $p_2 =0.39$) in continuous lines or by a node-based SPA which approximates the link-based system ($q_1 =0.68$ and $p_1 =0.56$ according to Eq. (\ref{transition})) in dashed lines. The two black lines perfectly overlap, while the node-based membership distribution is slightly shifted by the use of approximation (\ref{approx2}).}
  \label{s_compar}
\end{figure*}

\subsection{Scaling behavior}

From PA, it is well known that the $N^*_m$ and $S^*_n$ distributions will fall as power laws, i.e.,
\begin{equation}
N^*_m \propto m^{-\gamma _N} \;\;\; \textrm{and } \;\; S^*_n \propto n^{-\gamma _S} \; .
\end{equation}
To calculate the scaling exponent $\gamma _N$, we can evaluate the following ratio using Eq. (\ref{powers})
\begin{equation}
\lim _{m \rightarrow \infty} \frac{N^*_m}{N^*_{m-1}} = \lim _{m \rightarrow \infty} \left(\frac{m}{m-1}\right)^{-\gamma _N} = \lim _{m \rightarrow \infty} \frac{\left(m-1\right)\Gamma _s}{1 + m\Gamma _s}
\end{equation}
from which it follows that
\begin{equation}
\gamma _N = \lim _{m \rightarrow \infty} \dfrac{\log \left(\left(m-1\right)\Gamma _s \bigg / \left(1 + m\Gamma _s\right) \right)}{\log \left(\left(m-1\right)\bigg /m \right)} = 1 + \frac{1}{\Gamma _s} \; .
\label{scale1}
\end{equation}
Similarly, one can directly write for structures:
\begin{equation}
\gamma _S = 1 + \frac{1}{\Omega _s} \; .
\label{scale2}
\end{equation}

The node and community degree distributions, as compositions of two power-law distributions, will fall as the slower of the two original distributions. Noting that $\mathcal{N}'(x,t)$ and $\mathcal{S}'(x,t)$ will follow $\gamma _{N'} = \gamma _N - 1$ and $\gamma _{S'} = \gamma _S - 1$ because of the derivative, we obtain:
\begin{equation}
\gamma _D = \min \bigg\lbrace\gamma _N, \gamma _S - 1 \bigg\rbrace \;\;\; \textrm{and } \;\; \gamma _C = \min \bigg\lbrace\gamma _N - 1, \gamma _S  \bigg\rbrace \; .
\end{equation}

These results are validated on Fig. \ref{scaling}.

\section{Approximations and limitations \label{sec:approximations}}

To complete our description of the SPA process, this section examines some approximations that have either proven useful when reproducing empirical data with the SPA process or that correspond to limitations of the present formalism.

\subsection{Correspondence between system bases}

%\begin{figure}
%  \centering
%  \includegraphics[width=0.35\textwidth]{arXiv}
%  \caption{Community structure of the \emph{cond-mat arXiv} as measured by a link community algorithm \cite{ahn} (dots) and as modelled by link-based SPA %with $q=0.95$; $p=0.39$ in continuous lines or node-based SPA used to approximate a link-based system with $q=0.68$; $p=0.56$ according to Eq. (\ref{transition}) in dashed lines. The two black lines perfectly overlap, while one membership distribution is shifted by approximation (\ref{approx2}).}
%  \label{arXiv}
%\end{figure}

Some systems reproduced in \cite{SPA} with node-based SPA ($s=1$) are actually link-based, for example the author collaboration network of the \textit{cond-mat arXiv}, where authors only appear once they have at least one collaboration. The link between node and link-based SPA is done by ignoring structures of size one when compiling the final system.

In \cite{SPA}, we mention that the system base $s$ was not a parameter of the model per se, but depends on the information available or on the nature of the system. For instance, the World-Wide Web is mapped by following links between webpages, such that it is impossible to find a page with no links. The smallest structural unit is thus the link and not the webpage itself: it is a link-based system ($s=2$). Similarly, the author collaboration network of the \textit{cond-mat arXiv} is built through collaborations and thus excludes authors without any links. Despite this fact, it can modelled through node-based SPA by ignoring structures of size one at the very end of the process. Furthermore, structures of size one can rarely be detected in network data if they are not completely disconnected from the rest of the systems. Hence, it is useful to be able to ignore these structures at the end of the stochastic growth process, independently of the system base. 

For the size distribution, ignoring structures of size one simply implies a renormalization for structures of size two or greater. Noting the PGF for an approximate link-based SPA $\mathcal{S}^{\textrm{app}}_2(x)$ using the original node-based functions $\mathcal{S}_1(x)$, we can write:
\begin{equation}
\mathcal{S}^{\textrm{app}}_2(x) = \frac{\mathcal{S}_1(x) - S_1x}{\mathcal{S}_1(1) - S_1} \; .
\label{approx1}
\end{equation}
For the membership distribution, once again assuming homogeneous mixing, we must randomly remove the fraction of memberships which corresponds to the structures of size one. Using the composition of PGFs, this can be done by composing the membership PGF with the PGF for a binomial trial:
\begin{equation}
\mathcal{N}^{\textrm{app}}_2(x) = \dfrac{\mathcal{N}_1\left(x\left(1-\epsilon\right)+\epsilon\right)-\mathcal{N}_1\left(\epsilon\right)}{1-\mathcal{N}_1\left(\epsilon\right)}
\label{approx2}
\end{equation}
where $\mathcal{N}_1\left(\epsilon\right)$ corresponds to the elements left with no memberships and thus need to be removed from the system. This trial will remove a fraction $\epsilon$ of memberships, where $\epsilon$ corresponds to the fraction of memberships which are associated with structures of size one:
\begin{equation}
\epsilon = \frac{S_1}{\sum _n nS_n} = \frac{\mathcal{S}'_1(0)}{\mathcal{S}'_1(1)} \; .
\end{equation}
The validity of this approximate description and the effects of switching between system bases are illustrated on Fig. \ref{s_compar}. Note how changing the system base, while keeping the parameters constant, greatly modifies the produced system. This highlights both the validity of Eqs. (\ref{approx1}) and (\ref{approx2}) (which feature two levels of approximation of homogeneous mixing) and the importance of considering the influence of the system base on the scaling behavior.

To compare the results of approximated and actual link-based SPA for the same community structure, we first need to identify the relation between the parameter pairs $\lbrace q_1 , p_1 \rbrace$ and $\lbrace q_2 , p_2 \rbrace$ which is such that $\Gamma _1 = \Gamma _2$ and $\Omega _1 = \Omega _2$. From Eq. (\ref{gam_ome}), we obtain:
\begin{equation}
p_2 = \frac{p_1}{2-p_1} \;\;\; \textrm{and } \;\; q_2 = \frac{2q_1}{2-p_1} \; .
\label{transition}
\end{equation}

While it is easily verified that ignoring structures of size one in node-based SPA can result in statistical features similar to that of link-based SPA (see Fig. \ref{arXiv}), there exists one part\-icularly important structural difference between these two kinds of systems. Mainly, a true link-based system is necessarily fully connected as each new elements creates at least one link with the old elements, while node-based systems can create many disconnected components that may or may not end up interconnecting through new structures (depending on $q$ and $p$). In real link-based systems, there is no restriction on connectedness. For instance, the \textit{cond-mat arXiv} network of co-authors has one giant component which consists of $\sim 93\%$ of the system, but other smaller satellite components still exist. While both SPA versions illustrated on Fig. \ref{arXiv} create a similar community structure as the \textit{cond-mat arXiv}, the node-based version is actually closer to reality.

\begin{figure*}[]
  \centering
  \subfigure[]{\includegraphics[width=0.36\linewidth]{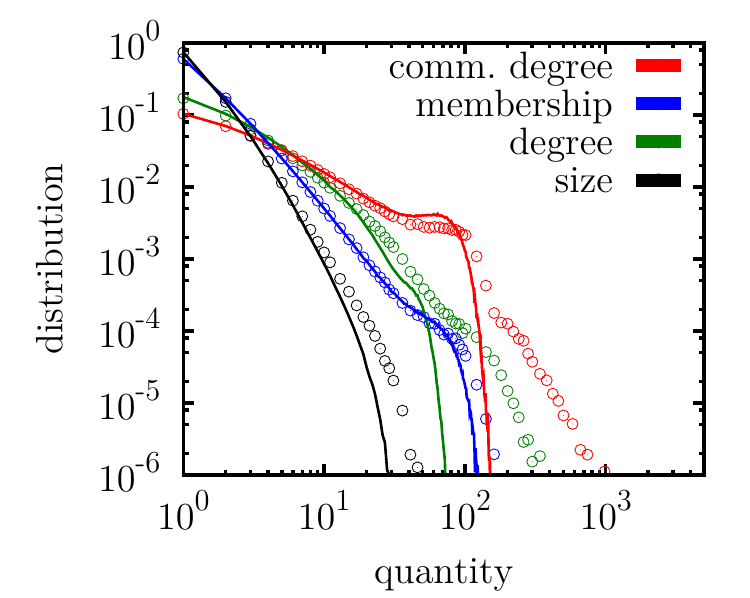} \label{multiple1}}
  \subfigure[]{\includegraphics[width=0.36\linewidth]{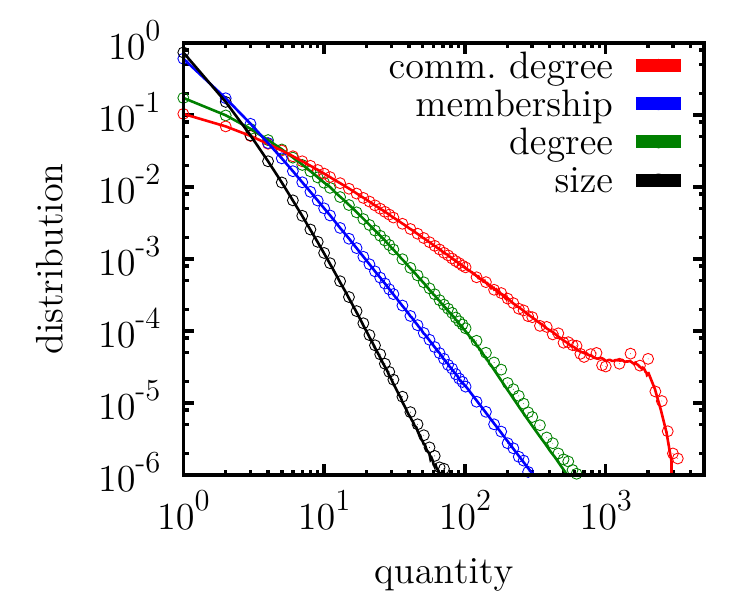} \label{multiple2}}
  \caption{(color online). (a) and (b) Comparison between the time evolution data presented in Fig. \ref{timeevo} (dots) and the same data when multiple memberships, multiple links and self-loops are discarded (lines) for systems with (a) 250 structures and (b) 25 000 structures. Multiple memberships, multiple links and self-loops are finite size effects which become negligible in the large-size limit.}
  \label{multiple}
\end{figure*}

\begin{figure}
  \centering
  \includegraphics[width=0.35\textwidth]{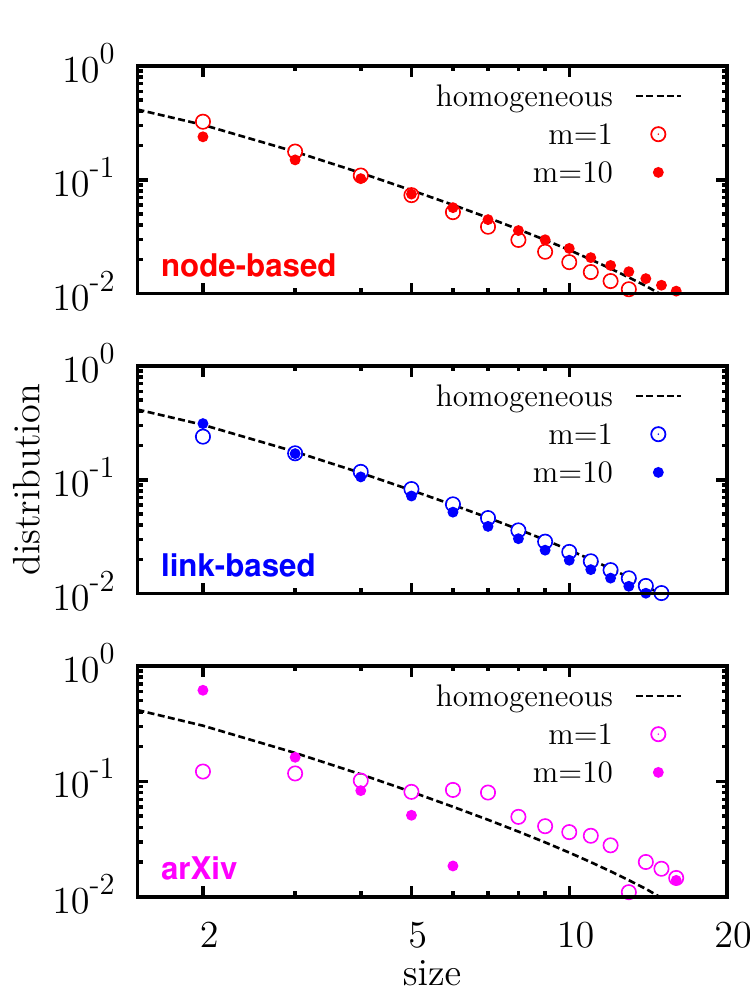}
  \caption{(color online). Size distribution of structures as seen from elements with different $m$ memberships. Markers represent empirical measures done on the \textit{cond-mat arXiv} and numerical results on the two SPA processes (using the parameters of Fig. \ref{arXiv}). The dashed line corresponds to what would be obtained through homogeneous pairing of memberships and structures.}
  \label{correlations}
\end{figure}

\begin{figure*}[]
  \centering
  \subfigure[]{\includegraphics[width=0.32\linewidth]{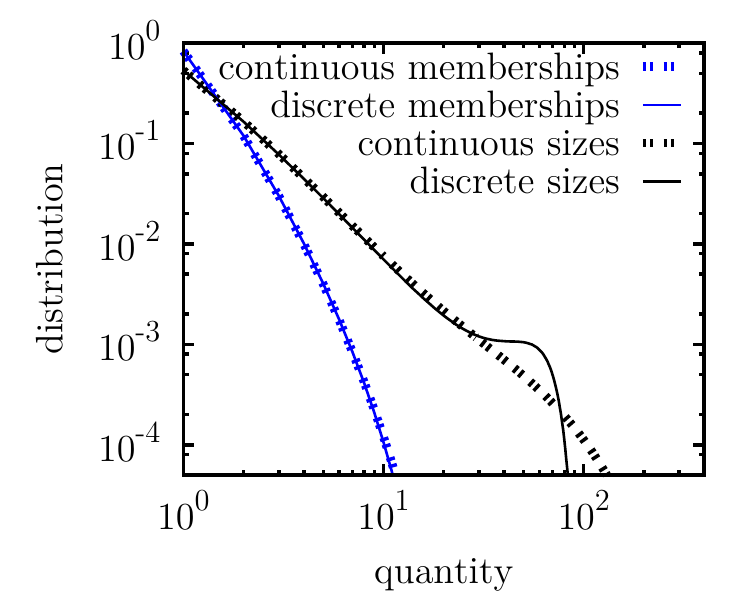} \label{DisVsCont}}
  \subfigure[]{\includegraphics[width=0.32\linewidth]{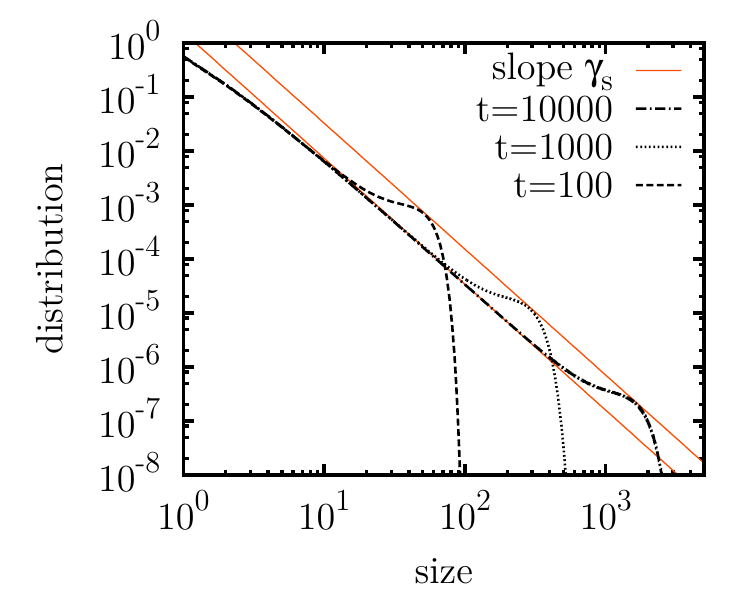} \label{decay}}
  \subfigure[]{\includegraphics[width=0.32\linewidth]{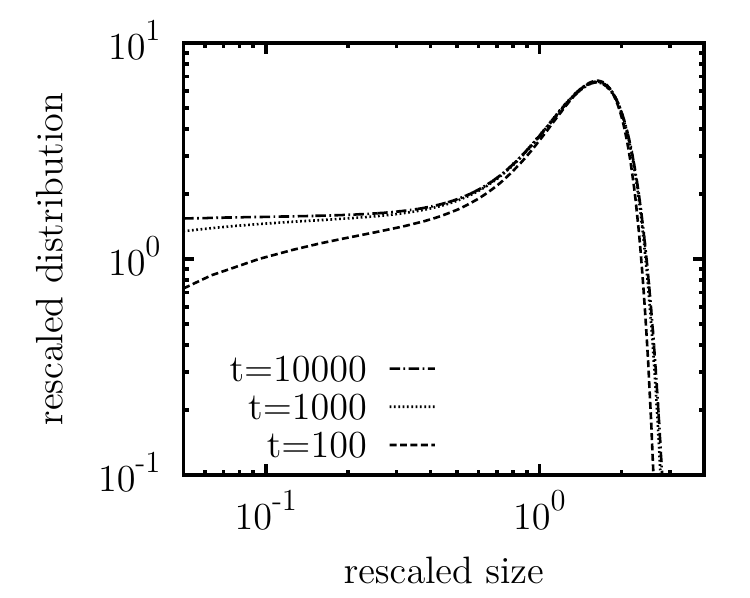} \label{superpose}}
  \caption{(color online). (a) Comparison of the memberships and sizes distributions of node-based SPA with $q=0.8$ and $p=0.2$ in discrete and continuous dynamics at time $t=100$. This illustrates how the peloton dynamics is a direct consequence of the maximal system size present only in the discrete version of the process. (b) The height of the peloton follows a power-law decay (here for the results of Fig. \ref{asympto2}) , such that its surface is conserved on a logarithmic scale as it evolves. The decay exponent of the peloton is the same as the scaling exponent of the distribution it creates. (c) Rescaled distribution $\{ n^{\gamma _s}S_n (t) \}$ as a function of rescaled community size $n/t^{1-p}$ highlights the scaling of the peloton dynamics.}
  \label{peloton}
\end{figure*}

\subsection{Multiple memberships, multiple links and self-loops}

In our description of the time evolution of SPA, we have never explicitly forbidden an element to join the same structure more than once. These multiple memberships, whose likelihood depends directly on the value of the $p$ or $q$ parameters, lead to multiple links between the same individuals and self-loops (where an element shares a structure with itself). Similarly, in our derivation of the degree distributions, we have supposed an infinite system where the probabilities that two structures overlap by more than one element fall to zero.

In empirical data, multiple links and self-loop are rarely considered. It can thus be useful to have an idea of the effect of such restrictions on SPA. Fig. \ref{multiple} presents two snapshots of the same scenarios of SPA, with or without forbidding multiple memberships, multiple links and self-loops when analyzing the final stage of the system. The cutoffs in the distributions of the first system are not surprising, as large and old structures are very likely to have recruited the same element more than once, especially with a small $q$. Yet, this effect rapidly becomes negligible as the system grows and we enter the large size limit in accordance with the assumptions of our analytical description (see Fig. \ref{multiple2}). 

%\begin{figure}
%  \centering
%  \includegraphics[width=0.35\textwidth]{Correlations}
%  \caption{Size distribution of structures as seen from elements with different $m$ memberships. Black markers represent empirical measures done on the \textit{cond-mat arXiv} using a link community algorithm \cite{ahn}; numerical averages are shown by red markers for node-based SPA simulations and blue markers for link-based SPA with the parameters of Fig. \ref{arXiv}. Differences between the node-based and link-based SPA processes are most likely due to the fact that the link-based version requires secondary founding elements for new structures, which are likely to be old elements.}
%  \label{correlations}
%\end{figure}

\subsection{Element-structure correlations \label{sec:correlations}}

Most of the approximations used throughout this paper are based on the assumption of homogeneous mixing: the elements belonging to a number $x$ of structures \emph{see} the same size distribution as the elements belonging to $y$ structures. This implies that there is no correlations except for the fact that an element is $x$ times more likely to belong to a given structure of size $x$ than to a particular structure of size one (\emph{natural correlations}). To investigate this matter, we compare the size distributions as seen from elements with different memberships in both the simulations done for Fig. \ref{arXiv} and the corresponding \emph{arXiv} data.

Figure \ref{correlations} presents the results of this investigation. First, the similitude between SPA and homogeneous mixing explains why our approximations were accurate. The small difference between the node-based and link-based SPA processes is most likely due to the fact that the link-based version requires more elements for the birth of new structures, which are consequently more likely to be old elements than in the node-based version. Second, there is a major difference between element-structure correlations in real-systems and SPA: elements with few memberships are much more likely to belong to larger structures in the arXiv data than in our SPA simulations. This shows how other levels of organization have yet to be taken into account in our stochastic models. Depending on what one wants to model, these correlations could potentially be important.

\begin{figure*}[]
  \centering
  \subfigure[]{\includegraphics[width=0.36\textwidth]{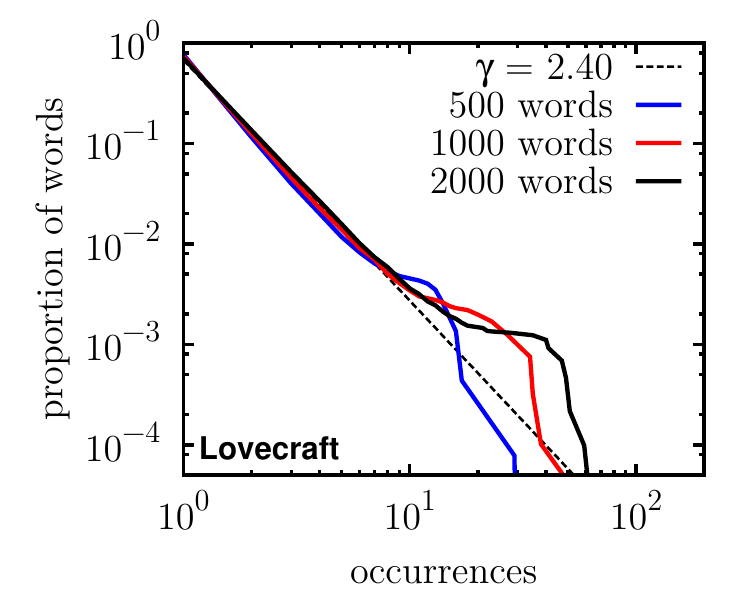}}
  \subfigure[]{\includegraphics[width=0.36\textwidth]{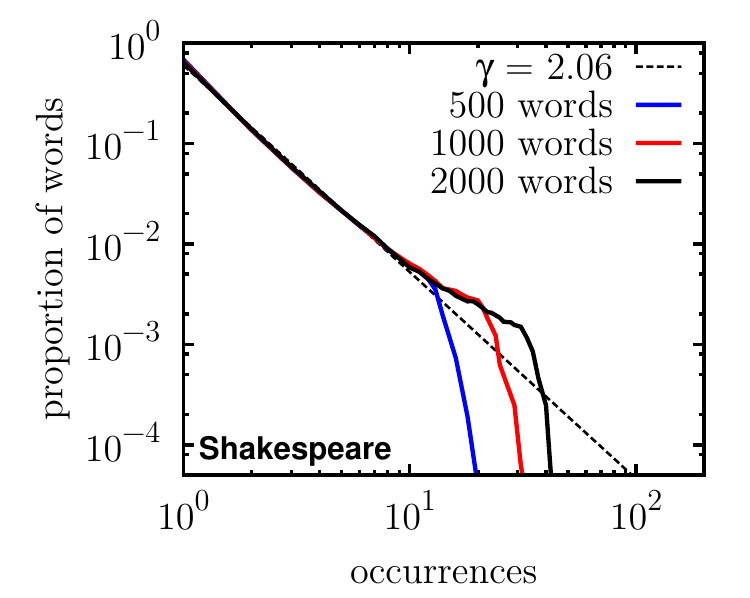}}\\
  \subfigure[]{\includegraphics[width=0.36\textwidth]{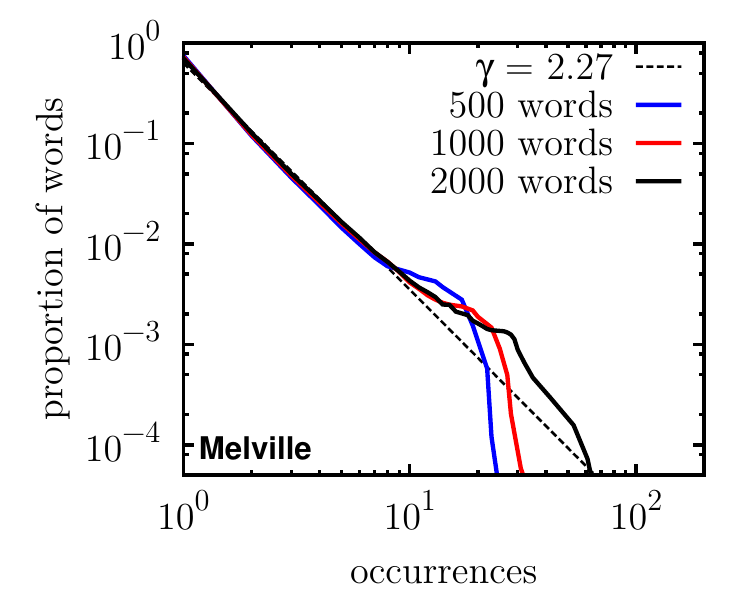}}
  \subfigure[]{\includegraphics[width=0.36\textwidth]{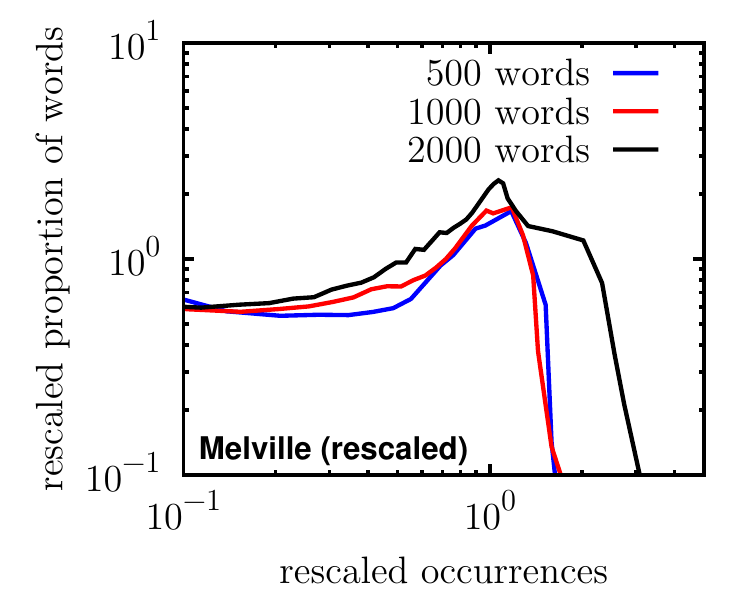} \label{rescaledmel}}
  \caption{(color online). Distributions of words by their number of occurrences in prose samples of different length taken from the complete works of (a) H.P. Lovecraft composed of nearly 800 000 words, (b) William Shakespeare with around 900 000 words and (c) Herman Melville with over 1 200 000 words. The peloton dynamics is manifest in all distributions. (d) The rescaling method of Fig. \ref{superpose}, with $\gamma = 2.27$ and $1-p = 0.43$, is applied to the statistics of Herman Melville's work.}
  \label{lovecraft}
\end{figure*}

\section{Peloton dynamics \label{sec:peloton}}

One particularly interesting feature of the results presented in Fig. \ref{timeevo} and \ref{asympto} is the dynamics of the entities in the tail of the distributions. In fact, these groups of individuals or structures resulted in clearly identifiable \emph{bulges} on their respective distributions. The dynamics of a system's leader is well-documented in the context of growing networks \cite{Krapivsky, Godreche} or word frequencies \cite{bernhardsson09}, but can be applied to any problem where one is interested in the statistics of the extremes (i.e., the growth of the biggest business firm, of the most popular website, etc.). What we observe here is that averaging over multiple realizations of the same experiment will result in the creation of a \emph{peloton} where one is significantly more likely to find entities than predicted by the asymptotic distribution (i.e., the leaders).

The clear distinction between the statistical distribution of leaders versus the rest of the system is a consequence of the maximal size of the system and of the limited growth resources available. To illustrate this claim, we can consider a continuous time version of PA in which there is no finite limitation to the number of growth events at every time step (see Appendix \ref{AA} for explicit solution of this process). Comparing the results of the discrete and continuous versions of our stochastic process on Fig. \ref{DisVsCont} illustrates how limiting growth resources results in the condensation of the leaders in a peloton. This draws a strong parallel between discrete preferential attachment and some sandpile models known to result in scale-free avalanche size distributions through \emph{self-organized criticality}. In some cases, such as the Oslo model (see \cite{christensen} \textsection 3.9), the biggest avalanches are limited by the size of the considered sandpile and are thus condensed in bulges identical to our pelotons.

Also striking is the fact that this peloton conserves its shape on a log-log scale (see Fig. \ref{decay}). To highlight this feature, Fig. \ref{superpose} rescales the distributions to account for the scaling in size ($\gamma _s$) and the peloton growth through time ($t^{1-p}$, see Appendix \ref{AB} for derivation). This rescaling method was borrowed from \cite{christensen} \textsection 3.9.8.

Leaders emerge in every single preferential growth realization, while the peloton dynamics can only manifest itself once we average over multiple systems or over many characteristic time scales of a single system (through the births and deaths of many different leaders). Consequently, empirical observations of this phenomenon are rare, because on the one hand we have only one Internet, one arXiv, and basically a unique copy of most complex systems, and on the other hand, we rarely have access to extensive data through long time scales. We can however find a solution if we go back to the first example used by Simon \cite{Simon55} to derive his model: the scale-free distribution of words by their number of occurrences in written text (i.e., Zipf's law \cite{zipf}). In this context, $q$ equals zero and the $p$ parameter corresponds to the probability that each new written word has never been used before. We can therefore consider different samples of text of equal length written by the same author as different realizations of the same experiment. 

With this in mind, we have picked different authors according to personal preferences and size of their body of work and divided their \oe{}uvres in samples of given lengths which we then used to evaluate Zipf's law under averaging (see Fig. \ref{lovecraft}). As predicted by PA, taking the average of multiple realizations of the same experiment results in a peloton which diverges from the traditional Zipf's law. In this case, the peloton implies that the leaders of this system (i.e., the most frequent words) consistently fall in the same scale of occurrences.

%It is noteworthy that a significant number of models which reproduce Zipf's law for written texts assume that it represents an equilibrium of the writing process (e.g., \cite{baek}). Yet, the peloton dynamics emerges from the time evolution of the system and thus from the fact that it is far from equilibrium. Therefore, the observation of peloton dynamics in written texts yields much more credibility to the corresponding stochastic growth principle.

Lastly, Fig. \ref{rescaledmel} reproduces the scaling analysis of Fig. \ref{superpose} for empirical results on prose samples. The varying surface of the peloton hints at a non-constant growth rate: a well-known feature of written text (see \cite{heap} \textsection 7.5).

\section{Conclusion}

In this paper, several analytical results for \emph{structural preferential attachment} have been obtained: solutions for its time evolution and asymptotic behavior as well as approximations for its different degree distributions. Those approximate descriptions are especially useful when it comes to using organization models as part of modelling efforts. 

We have also highlighted one particular shortcoming of the model: element-structure correlations. That is, SPA lacks any modelling or predictive power when it comes to asking \emph{who belongs to what structure}.

On the other hand, we have observed an interesting behavior of both the SPA and the classic PA models: \emph{the peloton dynamics}. This particular feature is important in order to predict the position of the leaders of a PA growth process. More interestingly, we have been able to observe this behavior in the growth of prose samples, which differentiates the PA principle from the other models generating scale-free designs but failing to predict this property.

The presentation of shortcomings and successes of the SPA principle (in terms of predictive value) shows the importance and the need for further study in stochastic growth models. 

\begin{acknowledgments}
The authors thank Yong-Yeol Ahn \textit{et al.} for their link community algorithm and Gergely Palla for providing the arXiv dataset. We also wish to acknowledge the help of Jean-Gabriel Young and Sebastian Bernhardsson for useful comments and criticism. The research team is grateful to NSERC, FQRNT and CIHR for financial support.
\end{acknowledgments}

\appendix

\section{Explicit solution to continuous time SPA \label{AA}}

Section \ref{sec:peloton} has presented an explicit solution for the time evolution of SPA in continuous time. This Appendix summarizes its derivation, based on a recently proposed  method \cite{morin}.

\subsection{Definition of a continuous time PA process}

The transition to continuous time simply implies that $q$ and $p$ now refer to birth rates for both elements and structures. The corresponding rates $1-q$ and $1-p$ thereby  correspond to the growth rates of existing elements and structures, respectively. This means that in a given time interval $[t, t+1]$, this new stochastic process could create an infinite number of elements with probability $\lim _{dt \rightarrow 0}\left(qdt\right)^{1/dt}$; whereas the discrete version could only create one element with probability $q$. While it is highly improbable that continuous time PA results in a system several orders of magnitude larger than $qt$ or $pt$, there is no maximal size per se.

This sort of continuous time dynamics is better described using simple ODEs, or master equations, as was done in \cite{SPA}. To this end, we once again follow $\tilde{N}_m$, the number of elements with $m$ memberships, and $\tilde{S}_n$, the number of structures enclosing $n$ elements. Using the same logic behind Eqs. (\ref{d1}) and (\ref{d2}), but considering infinitesimal time steps $dt$, one can write
\begin{align}
\tilde{N}_m(t+dt)\! = \tilde{N}_m(t)\! + dt\mathbf{\bigg\lbrace}\frac{\Gamma _s}{t}\left((m\! -\! 1)\tilde{N}_{m-1}(t)-\! m\tilde{N}_m(t)\right) +\! q \, \delta _{m1}\mathbf{\bigg\rbrace} \nonumber
\end{align}
and
\begin{equation}
\tilde{S}_n(t+dt)\! = \tilde{S}_n(t)\! + dt\mathbf{\bigg\lbrace} \frac{\Omega _s}{t}\left((n\! -\! 1)\tilde{S}_{n-1}(t) -\! n\tilde{S}_n(t)\right) +\! p \, \delta _{ns} \mathbf{\bigg\rbrace} \; , \nonumber
\end{equation}
which are straightforwardly rewritten as two ODEs:
\begin{equation}
\frac{d}{dt}\tilde{N}_m(t) = \frac{\Gamma _s}{t}\left((m-1)\tilde{N}_{m-1}(t)-m\tilde{N}_m(t)\right) + q\, \delta _{m1} \; ;
\label{node_master}
\end{equation}
\begin{equation}
\frac{d}{dt}\tilde{S}_n(t) = \frac{\Omega _s}{t}\left((n-1)\tilde{S}_{n-1}(t) - n\tilde{S}_n(t)\right) + p \, \delta _{ns} \; .
\label{struc_master}
\end{equation}
Because these two last equations have the same form, we solve them separately using a general continuous time PA equation. Consider
\begin{equation}
\frac{d}{dt}P_k(t) = \beta\, \delta _{km} + R_{k-1}(t)P_{k-1}(t) - R_k(t)P_k(t)
\label{gen_eq}
\end{equation}
where $\beta$ is the birth rate, $m$ is the size of new entities and $R_i(t)$ is the attachment rate on entities of size $i$, which we define using a growth rate $\alpha$, an initial total size $m_0$ and a normalization rate $\lambda$:
\begin{equation}
R_i(t) = \frac{\alpha i}{m_0 +\lambda t } \; .
\end{equation}
It proves useful to rewrite (\ref{gen_eq}) in dimensionless form as
\begin{equation}
 \frac{d}{d\tau}P_k(\tau) = \overline{\beta}\, \delta _{km} + \overline{R}_{k-1}(\tau) P_{k-1}(\tau) - \overline{R}_k(\tau) P_k(\tau)
\label{nodim_gen_eq}
\end{equation}
with dimensionless time $\tau = \alpha t$, parameters $\overline{\beta}= \beta/\alpha$, $\overline{\lambda}= \lambda/\alpha$, and attachment rate
$\overline{R}_k(\tau) = k / (m_0 + \overline{\lambda} \tau)$ respectively.    
Table $\ref{table}$ gives the values of the different parameters for the classical PA models and for SPA.

\begin{table}[h]
\begin{center}
\begin{tabular}{c|c|c|c|c|}
\cline{2-5}
& \multicolumn{2}{|c|}{PA} & \multicolumn{2}{|c|}{SPA}\\ \cline{2-5}
& Simon & BA & elements & structures \\ \cline{1-5} \cline{1-5}
\multicolumn{1}{|c|}{$\quad \beta/\alpha \quad$} & $\; p/(1-p) \;$ & $\quad 1/m \quad$ & $q/\alpha$ & $p/\alpha$ \\ \cline{1-5}
\multicolumn{1}{|c|}{$\alpha$} & $1-p$ & $m$ & $1-q+p(s-1)$ & $1-p$ \\ \cline{1-5}
\multicolumn{1}{|c|}{$ \lambda/\alpha$} & $1/(1-p)$ & $2$ & $[1+p(s-1)]/\alpha$ & $[1+p(s-1)]/\alpha$ \\ \cline{1-5}
\multicolumn{1}{|c|}{$m$} & $1$ & $m$ & $1$ & $s$ \\ \cline{1-5}
\end{tabular}
\end{center}
\caption{Parameters of the general PA process (Eq. \ref{nodim_gen_eq}) in the context of Simon's model \cite{Simon55}, of the Barab\'{a}si-Albert model (BA) \cite{barabasi99} and of SPA.}
\label{table}
\end{table}

\subsection{Explicit solution}
Let
\begin{equation}
\overline{H}_k(t) = \textrm{exp}\left[ \int \overline{R}_k(\tau) d\tau \right] = \left( m_0 + \overline{\lambda} \tau \right)^{k / \overline{\lambda}} \; ,
\label{H}
\end{equation}
so that Eq. (\ref{nodim_gen_eq}) can be written as:
\begin{equation}
\frac{d}{d\tau}\left[P_k(\tau) \overline{H}_k(\tau)\right] = \overline{\beta} \overline{H}_k(\tau) \delta _{km} 
                         + \overline{R}_{k-1}(\tau) \overline{H}_k(\tau) P_{k-1}(\tau) \; .
\end{equation}
The general solution of this transformed equation is:
\begin{eqnarray}
P_k(\tau) &=&  \overline{\beta} \frac{   (m_0 + \overline{\lambda} \tau)}{k + \overline{\lambda}} \delta_{km} \nonumber \\
               &+ &  \frac{(1 - \delta_{km})}{\overline{H}_k(\tau)}\int \overline{R}_{k-1}(\tau) \overline{H}_k(\tau) P_{k-1}(\tau) d\tau  +C_k \; ,
\end{eqnarray}
where $\{C_k\}$ are constants of integration determined by the initial conditions. 
Solving for the first few values of $k$ ($m$, $m+1$, $m+2$, \ldots) reveals the following pattern for the solutions:
%\begin{align}
%P_{m+k}(t) = & \frac{\overline{\beta} \prod _{j=0}^{k-1} \left(m + j\right)}{\left(m + \overline{\lambda}\right)\prod _{j=1}^{k}\left(m+\overline{\lambda} +j \right)} \left(m_0 +\lambda t \right) \nonumber \\
%& \quad + \sum _{i=0}^k\frac{C_{m+i}\prod _{j=i}^{k-1}\left(m+j\right)}{\left(k-i\right)!\left(m_0 + \lambda t \right)^{(m+i) / \overline{\lambda}}}
%\end{align}
\begin{eqnarray}
P_{m+k}(\tau) &= &  \overline{\beta} \frac{ (m)_k }{ (m+\overline{\lambda})_{k+1} }  \left( m_0 + \overline{\lambda} \tau \right) \nonumber \\
&+& \sum_{i=0}^k \frac{ (m)_k }{ (m)_i } \frac{ C_{m+i} }{\left( k-i \right)!}    
                                                   \left( m_0+ \overline{\lambda} \tau \right)^{-(m+i) / \overline{\lambda} } 
\label{gen_soln}
\end{eqnarray}
where $(\gamma)_j \equiv (\gamma) (\gamma +1) \ldots (\gamma +j -1)$ are Pochammer symbols.  The last step towards a 
complete solution is to determine an explicit form of the constants of integrations $\{ C_{m+k}\}$ in terms of the initial conditions 
$\{ P_{m+k}(0)\}$.
This is easily accomplished by writing (\ref{gen_soln}) in a matrix form for the vector of initial conditions $\boldsymbol{P}(0)$ 
\begin{equation}
      \boldsymbol{P}(0) = \boldsymbol{A}(0) + \bf{L}(0) \boldsymbol{C}
\end{equation}      
in terms of the vector $\boldsymbol{C}$ of integration constants and a {\em lower triangular} matrix $\bf L$, followed by the observation
that the inverse of a (lower/upper) triangular matrix is also a (lower/upper) triangular matrix whose elements can be constructed by forward substitution. Given that the elements of ${\bf L}(0)$ are 
\begin{equation}
L_{m+k,m+i}(0) = \binom{m+k-1}{m+i-1}\frac{1}{m_0^{m+i}}
 \end{equation}
we find that the elements of the inverse matrix, denoted $\bf M$, are simply
\begin{equation}
M_{m+k,m+i} = (-1)^{k-i} \binom{m+k-1}{m+i-1}m_0^{m+i} \; .
\end{equation}
Inserting this solution in (\ref{gen_soln}), we get
\begin{equation}
             \boldsymbol{P}(\tau) = [\boldsymbol{A}(\tau) - {\bf L}(\tau) {\bf M} \boldsymbol{A}(0) ]
                                                    + \bf{L}(\tau) {\bf M} \boldsymbol{P}(0) \ ,
\end{equation}
which nicely isolates the principal dynamics (the first 2 terms) from the initial conditions. Specifically, by imposing the usual initial conditions,
$P_{m+k}(0) = \delta_{k0}$, it is straightforward, albeit somewhat lengthy, to obtain a closed-form  expression for the complete dynamical elements as
\begin{eqnarray}
   P_{m+k}(\tau) &=& \overline{\beta} m_0 (m)_k 
                             \left[ \frac{1}{(m+\overline{\lambda})_{k+1}} X(\tau ) \right. \nonumber \\
         &-& \left. \frac{1}{(m+\overline{\lambda})} \frac{1}{\Gamma(k+1)} X(\tau)^m  F_k(X(\tau))\right]    \nonumber \\
                       &+& (m)_k \frac{1}{\Gamma(k+1)} X(\tau)^m (1 -X(\tau))^k
\end{eqnarray}
with $X(\tau) = m_0 / (m_0 + \overline{\lambda} \tau)$ and where 
$F_k(X)=  {\ }_2F_1(-k,m+\overline{\lambda}; m+\overline{\lambda} +1; X)\ $
represents a terminating hypergeometric series of degree $k$. One verifies that, by setting $\tau=0$ in the previous expression, one obtains
 $P_{m+k}(0) = \delta_{k0}$ as it should.

It can further be shown that the continuous and discrete time versions of PA converge toward the same asymptotic behavior.

\section{Scaling exponents in the peloton dynamics \label{AB}}

%\begin{figure}[t!]
%  \centering
%  \includegraphics[width=0.35\textwidth]{LHD_fig10}
%  \caption{(color online). Gaussian fit to a peloton of Fig. \ref{superpose}.}
%  \label{gauss}
%\end{figure}

It has been seen 
in Fig. \ref{superpose}, that the probability distribution $P(x;t)$ follows the scaling relation
\begin{equation}
    \widetilde{P}(x) \propto x^{\gamma} P(x/f(t); t\gg 1) \ ,
\end{equation}
where $\gamma$ is either equal to $\gamma_N$ for elements or $\gamma_S$ for structures.
This Appendix derives the growth function, $f(t)$, describing  the mean state of a single entity (e.g., its number of occurences or its size) at time $t$ within a system whose global growth is governed by PA. Once again, because we follow mean quantities, the process is deterministic.    

Without loss of generality, we suppose that only one entity is present at time $t=1$, such that always exactly $t$ events will have occured by time $t$. This simplifies the normalization of transition probability and we can thus write the effect of a general PA  step on a single entity as:
\begin{align}
f(t+1) = \left[\beta + \alpha\frac{t-f(t)}{t}\right]f(t)  + \alpha\frac{f(t)}{t}\bigg(f(t)+1\bigg) \; .
\end{align}
For the node-based cases, a further simplification arises, $\alpha + \beta = 1$,  yielding a recursive rule for the growth function $f(t)$:
\begin{equation}
f(t+1) = \left(1+\frac{\alpha}{t}\right)f(t) \; ,
\end{equation}
which directly fixes the derivative in the limit of large $t$:
\begin{equation}
\frac{d}{dt}f(t) = \frac{\alpha}{t}f(t) \; .
\label{Bdiff}
\end{equation}
The general solution to Eq. (\ref{Bdiff}) is:
\begin{equation}
f(t) = At^{\alpha} + B \; .
\end{equation}
For the original entity, $f(1) = 1$, which is destined to be the leader of this deterministic process, one obtains the following mean position at time $t$:
\begin{equation}
f(t) = t^{\alpha} \; .
\label{Blead}
\end{equation}
Equation (\ref{Blead}) dictates the evolution of the leader's position and thus fixes the renormalization used in Fig. \ref{superpose}. Once again, one can refer to Tab. \ref{table} for the values of $\alpha$ in different PA models.% The full peloton can then be described through a Gaussian approximation around this mean value (see Fig. \ref{gauss}).

%\bibliographystyle{apsrev}
%\bibliography{./Biblio.bib}

\end{document}